\documentclass[12pt,a4paper]{article}
\usepackage{amsmath,amssymb}
\usepackage{units}
\usepackage{graphicx}
\DeclareMathOperator{\tr}{tr}

\newcommand{\ee}{\mathrm{e}}
\newcommand{\dd}{\mathrm{d}}
\allowdisplaybreaks
\title{%
  Supersymmetric~extensions and dark~matter in models of
  warped~Higgsless electroweak~symmetry~breaking}
\author{%
  Alexander Knochel%
    \thanks{e-mail: \texttt{aknochel@physik.uni-wuerzburg.de}}\\
  Thorsten Ohl%
    \thanks{e-mail: \texttt{ohl@physik.uni-wuerzburg.de}}\\
  \hfil\\
    Institut f\"ur Theoretische Physik und Astrophysik\\
    Universit\"at W\"urzburg\\
    Am Hubland, 97074 W\"urzburg, Germany}

\begin{document}
\maketitle
\begin{abstract}
  We introduce a minimal supersymmetric extension of a higgsless model
  for electroweak symmetry breaking in a warped extra dimension.
  In contrast to the non supersymmetric version, our model naturally contains a
  candidate for cold dark matter.  No KK-parity is required, because its
  stability is guaranteed by an~$R$-parity.  We discuss
  the regions in parameter space that are compatible with the observed dark
  matter content of our universe and are allowed by electroweak
  precision measurements as well as direct searches.
\end{abstract}

\section{Introduction}

The electroweak standard model~(SM) has proven to be extremely
successful in describing the hitherto collected data from particle
physics experiments.  However, we are still waiting for the
forthcoming LHC experiments to find the first direct experimental
evidence for the detailed dynamics behind electroweak symmetry
breaking~(EWSB).  In the minimal SM, EWSB is accomplished by the
condensation of a fundamental scalar Higgs field, but the corresponding
particle has, so far, eluded detection at LEP and the Tevatron.
While the minimal~SM with a rather light Higgs particle describes
the electroweak precision data very well, any model with just a light
fundamental scalar is not natural when a much higher scale exists,
such as the grand unification and quantum gravity scales.  Therefore
the minimal SM is expected to be just the low energy effective theory
of an extended model in which new physics, that
protects the mass of the Higgs particles, appears at the TeV-scale,
e.\,g.~supersymmetry~(SUSY), little Higgs or extra dimensions.

Instead of adding particles to the minimal SM in order to protect the
naturalness of
a light fundamental scalar, one can instead attempt to break the
electroweak symmetry without invoking the condensation of fundamental
scalar fields.  Indeed, such models were already
proposed~\cite{Weinberg:1975gm} a few years after the introduction of
the SM.  In this class of models a new strong interaction, called
Technicolor~(TC), effects EWSB in analogy to chiral symmetry
breaking in QCD, but at the TEV scale.  Unfortunately, after
extending~TC in order to produce fermion masses (ETC), the resulting
models suffer from severe problems concerning flavor changing
currents and electroweak precision observables\footnote{Note, however,
that some progress has been made in recent
years~\cite{Christensen:2005cb,Christensen:2006rf}.}.
Recently a deeper understanding, based on the celebrated AdS/CFT duality,
of such higgsless models has emerged which is inspired by EWSB in
extra dimensions, particularly in warped spacetimes such as a slice of
AdS5~\cite{Csaki:2003dt,Csaki:2003zu}.  Even though these approaches
are nonrenormalizable and still suffer from some of the problems that
plague ETC, they remain perturbatively calculable up to energies
of~$5\,\ldots\,\unit[10]{TeV}$~\cite{Davoudiasl:2004pw} while
simultaneously improving the fits to electroweak precision
observables~\cite{Cacciapaglia:2004rb,Cacciapaglia:2006gp}.

Combining observations of galaxy clusters, type Ia supernovae and the
cosmic microwave background~(CMB) from COBE and WMAP, there is now
overwhelming evidence that nonbaryonic cold dark matter~(CDM)
constitutes roughly $20\%$ of the energy density of the universe~\cite{Spergel:2006hy}.
This CDM should predominantly consist of stable nonrelativistic,
electrically neutral colorless particles (WIMPs).  In light of this
situation, it seems inevitable for any extension or alternative to the
SM to eventually incorporate a phenomenologically
acceptable CDM candidate.

In SUSY models and in models with flat extra dimensions, there are
natural candidates for CDM that are stabilized by $R$- and
KK-parity~\cite{Servant:2002aq,Kong:2005hn} respectively.  However, the
warp factors and symmetry breaking boundary terms on the branes that
are characteristic of higgsless models of EWSB break
the translational invariance in the extra dimensions.  Therefore, the
corresponding KK-parity is not a symmetry of higgsless models.  As a
result, none of the Kaluza-Klein~(KK) modes are protected against
decay and they fail to provide a candidate for stable CDM.  In the light of the
above astrophysical observations, we are therefore compelled to extend
the spectrum of higgsless models in a different way if we want to
retain them as viable alternatives to the minimal supersymmetric
standard model~(MSSM) and its cousins.  One approach, which has been
proposed recently~\cite{Agashe:2007jb}, is to glue
together two slices of AdS5 and to use an exchange symmetry as a
KK-parity, protecting a CDM candidate from decay. Another
approach~\cite{Panico:2008bx} extends the spectrum and guarantees the
stability of a CDM candidate by a~$Z_2$ symmetry.

While the naturalness of the TeV-scale does not require the
introduction of SUSY in higgsless models, SUSY can nevertheless be
expected to play an important role in any more fundamental theory.  In
particular, string theory requires SUSY at some scale. Indeed, string
theory provided much of the inspiration for the revival of extra
dimensional models, for higgsless models and for AdS/CFT.  It is
therefore natural to investigate supersymmetric extensions of
higgsless models.  If these extensions admit an $R$-parity, stable
candidates for CDM are guaranteed to be part of the spectrum.  Indeed, we
propose to extend the above-mentioned higgsless models in a slice of
AdS5 with SUSY, providing a stable particle as a viable CDM candidate.

In this paper, we introduce the following framework: assume that
the sector providing EWSB exhibits a global SUSY whereas
the~$SU(2)_L\times U(1)_Y$ gauge theory to which it is coupled does
not. Our approach extends the particle spectrum and its interactions in a
well-defined way requiring a minimum of new assumptions. We examine
possible scenarios which differ in how the fields on the UV brane
are coupled to the bulk, calculate the relic density and show that the
observed dark matter density $\Omega h^2 \approx
0.09\,\ldots\,0.11$~\cite{Spergel:2006hy} can be achieved
straightforwardly with reasonable gaugino masses.

The paper is organized as follows: in section~\ref{sec:5dsusy} we fix
our notations, by giving a concise review of SUSY and the SUSY
spectrum in a warped background. In section~\ref{sec:model} we
outline the supersymmetric version of the higgsless models on which
our construction is
based~\cite{Csaki:2003dt,Csaki:2003zu,Cacciapaglia:2004rb,Cacciapaglia:2006gp}.
In section~\ref{sec:dmpheno} we survey the parameter space of our
proposed model, identifying the CDM candidate and estimating its relic
density.  We find that it is possible to obtain realistic relic
densities of a neutralino LSP without much fine tuning.
Section~\ref{sec:conclusions} summarizes our conclusions.  Further
technical details, regarding the interplay of SUSY, KK-decomposition
and boundary conditions~(BCs) can be found in appendix~\ref{appendix}.

\section{Global Supersymmetry in a Slice of AdS5}
\label{sec:5dsusy}

In this work we assume that the backreaction of the matter and gauge
degrees of freedom---including superpartners and KK-towers---on the
warped geometry can be neglected.  Therefore, we only need to
implement the global SUSY transformations that are compatible with the
isometries of the warped background geometry and not the full supergravity algebra.

In the case of flat extra dimensions, the SUSY algebra in Minkowski
space can be generalized straightforwardly to five dimensions by using
the corresponding
Clifford algebra and promoting the parameters to four component
spinors. The commutator of two transformations then still reads
\begin{equation}
\label{eq:5D/N=1}
   [\delta_\eta,\delta_\xi]
     = -2(\bar{\eta}\gamma^M\xi-\bar{\xi}\gamma^M\eta)P_M\,,
\end{equation}
where the gamma matrices are now defined such that
\begin{equation}
\label{eq:4D/N=2}
  \{ \gamma^M,\gamma^N \}=2 \eta^{MN},\qquad M,N=0\ldots3,5\,.
\end{equation}
Expressing the 5D~$N$=1-SUSY algebra in terms of four component generators,
\begin{equation}
  \{Q_i,\overline{Q}_j\}
    = -2\gamma^M_{ij} P_M =-2\gamma_{ij}^\mu P_\mu -2\gamma^5_{ij} P_5\,,
\end{equation}
a comparison with the 4D~$N$=2-SUSY algebra reveals that $iP_5$ plays
the role of a central charge in the 4D
picture~\cite{Sohnius:1985qm}.
Since translations
leave the metric~$\eta_{MN}$ invariant, we are dealing with a
global spacetime symmetry.

Moving on to a warped background, we retain the approximation of
neglecting the backreaction, as stated above.
Following the approach of~\cite{Hall:2003yc} for the
treatment of a curved background, we define global SUSY
transformations by demanding that they leave the
metric~$g$ invariant.  In other words, we demand that these global SUSY
transformations close into a Killing vector field~$v$ of the
background metric.
In particular
\begin{equation}
\label{eq:killing-1}
  [\delta_\xi, \delta_\eta]=v^N P_N
\end{equation}
and the Killing condition for~$v$ reads
\begin{equation}
\label{eq:killing-2}
  v^M \partial_M g_{AB}+g_{AM}\partial_B v^M+g_{BM}\partial_A v^M=0\,,
\end{equation}
while the gamma matrices satisfy now
\begin{equation}
  \{ \gamma^M,\gamma^N \}=2 g^{MN}\,.
\end{equation}
This gives us a condition for the spinor valued parameters of the SUSY
transformations, and solutions~$(\xi,\eta)$ of~(\ref{eq:killing-1})
and~(\ref{eq:killing-2}) are called Killing spinors.

We use a ``mostly~$-$'' metric convention. The warped background
metric we are dealing
with in this work is that of Randall-Sundrum~(RS)
Type~I~\cite{Randall:1999ee}, which, in proper distance coordinates, reads
$g_{\mu\nu}=\ee^{-2 R k y}\eta_{\mu\nu},~g_{55}=-R^2$, and
consequently~$\sqrt{g}=R
\ee^{-4 R k y}$. Here $\eta_{\mu\nu}$ denotes the Minkowski metric,
$R$ the radius, $k$ the RS curvature, and $y\in[0,\pi]$. As customary,
the 4D
spacetime at $y=0$\,($y=\pi$) will be referred to as the UV (IR)
brane. We work in the interval picture with BCs,
noting that this space could also be interpreted as an
orbifold. There are the usual 4D Poincar\'e symmetries and an
additional scaling symmetry $x^M \longrightarrow(1+\delta) x^M$
which is broken only by the presence of the branes. Working out
the Killing condition~(\ref{eq:killing-1})
and~(\ref{eq:killing-2}) in this background, one ends
up with a set of SUSY parameters which generate SUSY transformations
that close into the remaining symmetries, namely, using 2-spinor notation,
\begin{equation}
\label{eq:susypar}
   \xi(x,y)=\ee^{-R k y/2}\begin{pmatrix}\xi_\alpha^0\\0\end{pmatrix}\,,
\end{equation}
where the space-time dependence remains confined to the warp factor. This
relation fixes the KK wavefunctions of the superpartners.
Since~(\ref{eq:susypar}) is parameterized by a single Weyl-spinor,
there can be at most one 4D supersymmetry
left after integrating out the extra dimension.  Nevertheless, we
will see in the following sections, that the spectrum of the massive KK
modes will formally be that of 4D $N$=2-SUSY.

Counting the degrees of freedom reveals that a massless 5D vector boson
can not be combined with 5D spinors to form a SUSY multiplet. The
smallest multiplet that contains both is actually a
dimensionally reduced 6D vector multiplet which consists of a 4D
vector multiplet and a chiral multiplet~\cite{ArkaniHamed:2001tb}.

On the other hand, the 5D hypermultiplet is constructed from a
Dirac spinor $\Psi=(\psi_\alpha,\overline{\psi}^{c\dot\alpha})^T$
and two complex scalars as its superpartners. This makes it
equivalent to a combination of one 4D~chiral and one 4D~antichiral
multiplet. These relations have been extended to a full 4D
$N$=1-superfield formulation of these
multiplets~\cite{ArkaniHamed:2001tb,Marti:2001iw}, where the 4D kinetic 
terms arise from the superfield itself as usual while the 5D
dynamics are put in explicitly. Following~\cite{Marti:2001iw},
the superfields take the form
\begin{subequations}
\begin{align}
    V^a &= -\theta \sigma^m \overline{\theta} A_m^a-i \overline{\theta}^2\theta \lambda_1^a+i \theta^2
            \overline{\theta} \overline{\lambda}^a_1+\frac{1}{2}\theta\theta\overline{\theta}\overline{\theta}\, D^a\\
  \chi^a&= \frac{1}{\sqrt{2}}(\Sigma^a+iA_5^a)+\frac{i}{\sqrt{2}}\theta
             \sigma^m \overline{\theta} \partial_m( \Sigma^a+iA_5^a)-\frac{1}{4\sqrt{2}}
               \theta\theta\overline{\theta}\overline{\theta}\Box(\Sigma^a+iA_5^a) \nonumber\\
        &\hphantom{=}+\sqrt{2}\theta\lambda_2^a
           -\frac{i}{\sqrt{2}}\theta\theta\partial_m\lambda_2^a\sigma^m\overline{\theta}+\theta\theta C^a
\end{align}
\end{subequations}
for the gauge multiplet in Wess-Zumino gauge with the adjoint index~$a$ and
\begin{equation}
  H = h+i\theta \sigma^m \overline{\theta}\partial_m h -\frac{1}{4}\theta\theta\overline{\theta}\overline{\theta}\Box h
         + \sqrt{2}\theta \psi-\frac{i}{\sqrt{2}}\theta\theta\partial_m\psi
             \sigma^m \overline{\theta} +\theta\theta F
\end{equation}
for the chiral superfields~$H$ and~$H^c$ in the hypermultiplet.
In this notation, the action of the gauge multiplet for the
nonabelian case is given by
\begin{multline}
\label{eqgaugeaction}
   S_g[V,\chi] =\int\!\dd^5x
        \int\!\dd^2\theta\,\frac{R}{4N g^2}\tr[ W^\alpha W_\alpha]+\text{h.c.}\\
     + \int\! \dd^5x \int\!\dd^2\theta\, \frac{\ee^{-2Rky}}{R \,N g^2}
         \tr[(\sqrt{2}\partial_y+\overline{\chi})\ee^{-V}(-\sqrt{2}\partial_y+\chi)\ee^V
              +\partial_y\ee^{-V}\partial_y \ee^V ]\,,
\end{multline}
where $V=V^a T^a$ and $\chi=\chi^a T^a$ are the Lie-Algebra valued
superfields with normalization $\tr(T^aT^b)=N\delta^{ab}$ and $g/2$ is the 5D
coupling constant. The bulk action of the hypermultiplet coupled to
the gauge fields reads
\begin{multline}
\label{eq:hyperaction}
S_h [H,H^c,V,\chi]
   = \int\!\dd^5x \int\!\dd^2\theta\,  R \ee^{-2 R k y} \Bigg[\overline{H}
      \ee^{-V}H+H^c \ee^{V}{\overline H^c}\Bigg]\\
  +\int\! \dd^5x \int\!\dd^2\theta\, \ee^{-3 R k y}H^c
    \Bigg[\partial_y-\frac{1}{\sqrt{2}}\chi-(\frac{3}{2}-c)R k\Bigg]H+\text{h.c.}\,.
\end{multline}
Note that~$\partial_y$, $\chi$ and~$V$ are dimensionless whereas~$H$
and~$g$ have mass dimension~$3/2$ and~$-1/2$, respectively.  The
dimensionless quantity~$c$ is the 5D mass of the multiplet in units of
the RS curvature~$k$.  After the field
redefinitions described in appendix~\ref{sec:gaugebckk}, we obtain canonical
kinetic terms.  With our conventions, the gauge couplings in this
action correspond to the covariant derivatives
\begin{align}
D_M \Psi&=(\nabla_M-i A_M^a T^a)\Psi\\
(D_M \Phi)^a&=\nabla_M \Phi^a+ f^{abc}A_M^b \Phi^c\,.
\end{align}
Using these ingredients, we can now go on to construct the
supersymmetric model.

\section{The Model}
\label{sec:model}

\subsection{Higgsless Electroweak Symmetry Breaking}

Let us first give a short description of the
Higgsless models our construction is based upon
\cite{Csaki:2003zu,Csaki:2003sh}. They feature a left-right
symmetric gauge group\footnote{The general case~$g_L\neq g_R$ has been
studied in the literature and it turned out not to be an effective
means to improve precision fits and perturbative unitarity. We
therefore assume the gauge action to be LR symmetric in the bulk for
simplicity.}
\begin{equation}
G=SU(3)_C\times SU(2)_L\times SU(2)_R \times U(1)_X
\end{equation}
where $X$ will be the $(B-L)/2$ quantum number which is $1/6$ for
quarks and $-1/2$ for leptons. The corresponding coupling
constants are~$g_{5s}$, $g_L=g_R=g_5$ and~$\tilde{g}_5$.
The symmetry breaking by BCs is designed such that
\begin{equation}G\rightarrow
\begin{cases}
 SU(3)_C\times SU(2)_L\times U(1)_Y\quad \text{on the UV brane}\\
 SU(3)_C\times SU(2)_D\times U(1)_X \quad \text{on the IR brane}
\end{cases}\,,
\end{equation}
leaving only an overall $SU(3)_C\times U(1)_{EM}$ unbroken. The
diagonal subgroup $SU(2)_D$ generated by $T_D^a=T_L^a+T_R^a$ acts as
custodial symmetry. The BCs for the gauge fields (and later on
those of their scalar and fermionic superpartners) must be compatible with
this breaking scenario and with the vanishing of the variation of
the boundary action.
This is discussed in appendix~\ref{sec:gaugebckk}
together with the resulting KK decomposition. By
virtue of these BCs, the $W^\pm$ is a mixture of the
$SU(2)_L$ and $SU(2)_R$ gauge bosons (localized in the
$SU(2)_L$ field up to one permille in order to ensure the observed $V-A$
coupling), the $Z$ and $\gamma$ are a mixture of the $SU(2)_L$,
$SU(2)_R$ and $U(1)$ gauge bosons. The masses of $W^\pm$ and $Z$
are given approximately by\footnote{%
If the ratio~$g_5/\tilde{g_5}$ is determined from the particle masses, the deviations
from the SM are shifted to the couplings. To make the tree level corrections ``oblique'',
this quantity should be defined by fixing the gauge couplings to
matter first~\cite{Cacciapaglia:2004rb}.}
\begin{subequations}
\begin{align}
  m_W^2 &\simeq \frac{k \ee^{-2 R k \pi}}{(1+\kappa) R \pi} \\
  m_Z^2 &\simeq \frac{(g_5/\tilde{g}_5)^2 + \kappa +2}
                     {(g_5/\tilde{g}_5)^2 +1}\,m_W^2\,,
\end{align}
\end{subequations}
(where the brane kinetic term~(\ref{su2bkt}) contributes the
$\kappa$-dependence) so the radius~$R$ is determined by the RS
curvature~$k$ and the~$W$ mass.

The leptons and quarks are implemented as
in~\cite{Cacciapaglia:2004rb}.
There are two doublets transforming under $SU(2)_L$ and $SU(2)_R$
respectively for each SM fermion
\begin{subequations}
\begin{align}
  \Psi_L &= (\psi^u_L,\overline{\psi^{uc}_L},\psi^d_L,\overline{\psi^{d c}_L})^T\\
  \Psi_R &= (\psi^u_R,\overline{\psi^{uc}_R},\psi^d_R,\overline{\psi^{d c}_R})^T\,.
\end{align}
\end{subequations}
The two doublets get 5D Dirac masses denoted by $c_L$ and $c_R$ respectively 
(cf.~(\ref{eq:hyperaction})) which are allowed in the bulk where the theory is vectorlike.
In the
limit of massless fermion zero modes, we impose BCs such that $\psi_L$ and
$\psi^c_R$ have a zero mode for which $\psi_R$ and $\psi^c_L$
vanish. The mass of the zero mode is then
lifted by a $SU(2)_D$-invariant Dirac mass $\mu$ on the IR brane,
resulting in modified effective BCs which mix
$\Psi_L$ and $\Psi_R$.
So far the treatment is the same for the quarks and leptons. They
differ however in the way how the doublets are split on the UV
brane where the theory is not invariant under $SU(2)_D$. Quarks are
given a UV brane kinetic term with parameter $\rho$, whereas the
$SU(2)_R$ neutrinos can receive a large UV localized Majorana mass
$\mu_r$ which leads to a seesaw-like mechanism. The mass of each
fermion is determined by the UV splitting parameter, the IR mass and
the 5D ``bulk'' masses of the doublets, $c_L$ and $c_R$, which
control the localization of the fermion zero modes.
The resulting
masses are approximately (for $c_L>1/2$, $c_R<-1/2$)
\begin{subequations}
\begin{align}
  m_f &\simeq \frac{{\sqrt{\left( 1 - 2\,c_L \right) \,\left( 1 + 2\,c_R \right) }}\,
                   {\left( \ee^{R k \pi} \right) }^{ c_L - c_R-1}\,\mu }{{\sqrt{1 -
                   \left( 1 + 2 c_R \right) \,k\,{\rho }^2}}}\\
  m_\nu &\simeq \frac{\mu^2}{\mu_r}(2 c_L -1) \left(\ee^{- 2 R k \pi}\right)^{c_L-c_R-1}\,.
\end{align}
\end{subequations}

\subsection{The Supersymmetric Extension}

\begin{figure}
  \begin{center}
    \includegraphics[width=0.7\linewidth]{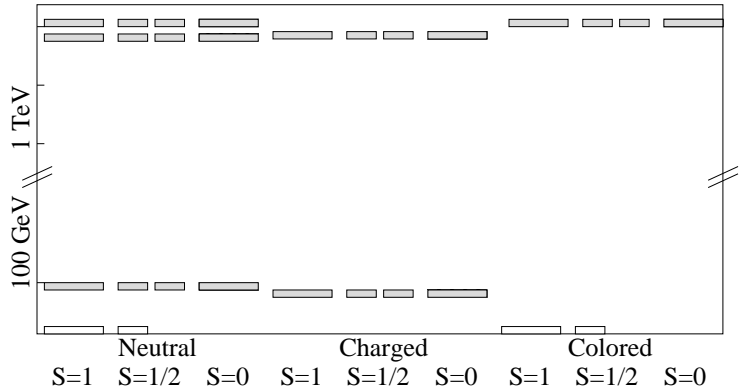}
  \end{center}
  \caption{\label{figsusyspec}%
    The particle spectrum with unbroken SUSY for a typical set of parameters
    $R k \pi\simeq$ $37.5$ ($k\simeq\Lambda_{\text{Pl}}$). The $S=1/2$ bars
    each represent one Majorana particle in the neutral column and one
    Dirac particle in the charged column, while the filled/unfilled $S=1$ bars distinguish
    between massive/massless vectors with three/two physical degrees of freedom}
\end{figure}

The structure of the 5D multiplets outlined above
implies that we must promote all fields to superfields separately
\begin{subequations}
\begin{align}
  \begin{pmatrix} A_\mu \\ A_5 \end{pmatrix} &\to
  \begin{pmatrix} V \\ \chi \end{pmatrix} \\
  \psi &\to H\\
  \psi^c &\to H^c\,.
\end{align}
\end{subequations}
Now
we can write the supersymmetric BCs for the gauge
multiplets in a short form on the IR brane (i.\,e.~$y=\pi$)
\begin{subequations}
\label{superbcs}
\begin{align}
  \begin{bmatrix} 1 & -1  \\ \partial_y & \partial_y  \end{bmatrix}
  \left.\begin{bmatrix} V^L \\ V^R \end{bmatrix}\right|_{y=\pi}
  =
  \left.\begin{bmatrix}
    \partial_y & -\partial_y \\
    1 & 1
  \end{bmatrix} \ee^{-2 R k y}
  \begin{bmatrix} \chi^L \\ \chi^R \end{bmatrix}\right|_{y=\pi} &= 0\,,\\
  \partial_y  V^X(\pi)
     = \chi^X(\pi)
     = \partial_y V^C(\pi)
     = \chi^C(\pi)
    &= 0
\end{align}
and on the UV brane (i.\,e.~$y=0$)
\begin{align}
  \left.\begin{bmatrix}
    \tilde{g}_5\partial_y & g_5 \partial_y  \\
    -g_5 & \tilde{g}_5
  \end{bmatrix}
  \begin{bmatrix} V^{R3} \\ V^X \end{bmatrix} \right|_{y=0} 
  =
  \left.\begin{bmatrix}
    \tilde{g}_5 & g_5   \\
    -g_5 \partial_y & \tilde{g}_5 \partial_y \\
  \end{bmatrix}
  \ee^{-2 R k y}
  \begin{bmatrix} \chi^{R3} \\ \chi^X \end{bmatrix} \right|_{y=0} &= 0\,, \\
  \partial_y V^L(0)
     = V^{R12}(0)
     = \partial_y V^C(0)
    &= 0\\
  \chi^L(0)
     = \partial_y \ee^{-2 R k y} \chi^{R12}(0)
     = \chi^C(0)
    &= 0\,.
\end{align}
\end{subequations}
These BCs are satisfied by solutions to the
Euler-Lagrange equations derived from the free part
of~(\ref{eqgaugeaction}). The BCs of the canonically
normalized fields are obtained after performing the redefinitions
(\ref{redef}) in appendix~\ref{sec:gaugebckk}. As it has already been
noted by the authors of~\cite{Marti:2001iw,Gherghetta:2000kr}, the
tree level spectrum remains highly degenerate even though one of the
supersymmetries has been broken by the warped background (shown for
our case in Fig.~\ref{figsusyspec}). Similarly, the generation of 
matter fermion masses on the IR brane is made supersymmetric by giving
$SU(2)_D$ invariant localized masses to the entire multiplet $\delta(\pi-y) (H^c_L H_R+H^c_R H_L)$. 
Like for the case with fermions only, there are various prescriptions to convert this into effective BCs, 
and the hypermultiplet becomes degenerate for unbroken SUSY.

Of course, similarly to the MSSM, such a spectrum is not acceptable
because massless neutralinos, gluinos and light charginos
below~\unit[95]{GeV} have already been ruled out by
experimental searches. Also, the gauge scalars are even under~$R$
parity and could be produced below the pair production
threshold. We will therefore
investigate the consequences of having no supersymmetry at all
on the UV brane. Starting from this premise, it is natural to remove
all scalars from the UV brane by imposing the BCs
\begin{equation}
   \Sigma^L(0)=\Sigma^R(0)=\Sigma^X(0)=\Sigma^C(0)=h_L^i(0)=h_{R}^{ic}(0)=0\,.
\end{equation}
This pushes the~$SU(3)$ and $U(1)_X$ gauge scalars up to a
mass\footnote{This is accurate if further localized kinetic terms are
absent. The values are given here for $k=\Lambda_{\text{Pl}}$.}
\begin{equation}
   m_{\Sigma^C}=m_{\Sigma^X}=z_0 k \ee^{-R k \pi}
      \simeq z_0 \sqrt{R k \pi(1+\kappa)}\,m_W\simeq\unit[1.2]{TeV}
\end{equation}
where $z_0\simeq 2.41$ is the first zero of the Bessel function~$J_0(z)$.
The sfermions lie in a similar
mass range, depending on the localization parameter $c$ of the
multiplet. For a massless fermion we can approximate the tree level
sfermion masses
\begin{align}
m_{\tilde{f}}&=z_0 k \ee^{-R k \pi}\,,
\end{align}
where, if $c_L\neq 1/2,c_R\neq -1/2$, $z_0$  is the first positive
root of $J_{c_L-1/2}$ for $\tilde{f}_L$,  $J_{|c_R+1/2|}$ for
$\tilde{f}_R$,
 $J_{|c_L-1/2|}$ for $\tilde{f}_L^c$ and  $J_{-c_R-1/2}$ for
 $\tilde{f}_R^c$.
The situation is different for the scalars from gauge groups that are
broken on the IR brane. These scalars receive smaller tree level
masses
\begin{equation}
m_{\Sigma^0}=m_{\Sigma^+}\simeq\sqrt{\frac{2 k}{R \pi}} \ee^{-R k
\pi}=\sqrt{2(1+\kappa)}\,m_W\,,
\end{equation}
which is very interesting from a phenomenological point of view.
However, their tree level coupling to fermions is of the form $\Sigma
\psi \psi^c$ and therefore suppressed with the fermion mass,
vanishing altogether in the massless fermion limit where it is
forbidden by the chiral symmetry. For a large range of parameters,
the coupling to leptons and quarks is similar to the corresponding
SM Higgs coupling. Note however that since it is not the vacuum
expectation value of this
scalar triplet that breaks electroweak symmetry, this
similarity does not extend to the coupling to gauge
bosons. Consequently, the
processes corresponding to
Higgsstrahlung and vector boson fusion in the SM are absent in our
model at tree level.  This provides an experimental signature for
distinguishing our model from the MSSM at the upcoming LHC experiments.

Now let us turn to the gauginos. We split the
gauginos and the gauge bosons with the localized kinetic
term~(\ref{su2bkt})\footnote{If one does not mind tachyonic
solutions above the cutoff of the effective theory at $\simeq10$
TeV, it is also possible to raise the chargino mass
sufficiently with a
localized kinetic term, but this possibility will not be
investigated further in this work.}.
We will study the following two sets of BCs which
project out all massless modes, while
maintaining the~$SU(2)_L\times U(1)_Y$ invariance (but not the full supergauge
invariance) and the stationarity of the action on the UV brane.  In
the first scenario
\begin{subequations}
\begin{equation}
\label{eq:scenario-1}
  \lambda_2^{L}= \lambda_1^{R}= \lambda_1^{X}=0\,,
\end{equation}
while in the second
\begin{equation}
\label{eq:scenario-2}
\begin{aligned}
  \lambda^{L}_1
   &= \lambda^{R12}_2 =0 \\
      \cos(\theta) \lambda^X_1+\sin(\theta) \lambda^{R3}_1
   &= \cos(\theta) \lambda^{R3}_2-\sin(\theta) \lambda^X_2
    = 0\,.
\end{aligned}
\end{equation}
\end{subequations}
In both cases, the lightest charginos will receive a tree level mass
\begin{equation}
  m_{\chi^+} \simeq \sqrt{1+\kappa}\, m_W\,,
\end{equation}
which requires
$\kappa\geq0.4$ to escape the current detection bounds. The resulting
mass of the lightest neutralino mode will then be
\begin{subequations}
\begin{equation}
  m_{\chi^0}=m_{\chi^+}
\end{equation}
in the first scenario~(\ref{eq:scenario-1}) and
\begin{equation}
  m_{\chi^0}\simeq\cos(\theta) \, m_{\chi^+}
\end{equation}
\end{subequations}
in the second~(\ref{eq:scenario-2}).

\begin{figure}
\begin{center}
\includegraphics[width=.6\linewidth]{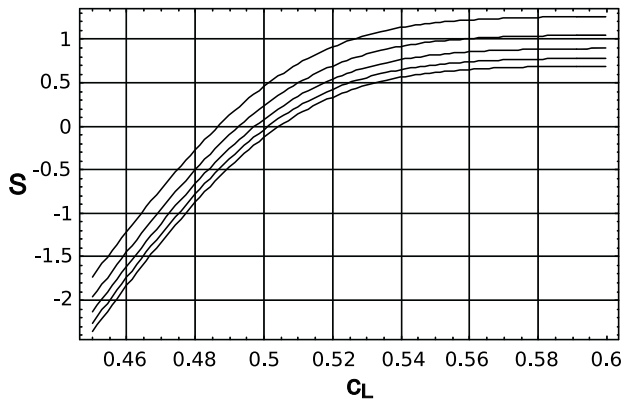}
\caption{The tree level contribution to the S parameter for a
massless probe fermion, $R k \pi\simeq 37.5$, $k\simeq \Lambda_{\text{Pl}}$
, $c_R=-c_L$ and different values of $\kappa$.
($\kappa=0,0.2,0.4,0.6,0.8$ from highest to
lowest)\label{spar}}\end{center}
\end{figure}

If we isolate the oblique corrections that arise at tree level
analogous to \cite{Cacciapaglia:2004jz,Cacciapaglia:2004rb},
there are two parameters that have a strong influence on the
parameters $S$, $T$ and $U$ in our scenarios, namely $\kappa$ and the
fermion localization parameters~$c$. Using the method of these authors, the
resulting value for $S$ is given in fig. \ref{spar}. We find a
combination of the results in \cite{Cacciapaglia:2004rb} and
\cite{Cacciapaglia:2004jz}. The kinetic term used to split the
electroweak gauge multiplets moves $S$ in the right direction and
can allow for delocalized or slightly UV localized fermions.

\subsection{Supergravity}

In some supersymmetric extensions of the SM, the gravitino poses a
problem for cosmological observables, either because it is overabundant
or so long-lived that its decay products spoil nucleosynthesis. So
far we have not included supergravity in the bulk, and we want to
give an argument that the inclusion of KK gravitons and gravitinos
in the bulk is possible without loosing the desired properties of
the model. A holographic interpretation along the AdS/CFT
correspondence where the energy momentum tensor and the
supersymmetric currents are present on the CFT side also warrants
the inclusion of supersymmetric gravity multiplets. Similarly to BC
gauge symmetry breaking we adopt the philosophy that the simple
picture with BCs is a description mimicking a full
theory with symmetry breaking in a slice of AdS5
\cite{Hall:2003yc,Gherghetta:2002nr}. By setting the
appropriate BCs we obtain a heavy short-lived gravitino.

Solving and KK expanding the (free) Rarita-Schwinger equations with
``twisted'' IR BCs
$\psi_2^\mu(0)=\psi_1^\mu(\pi)=0$, we obtain
a very light gravitino $m_0\simeq\sqrt{8}k\ee^{-2 R k
\pi}$\cite{Gherghetta:2000kr}. However, twisting the BC
on the UV brane yields the condition
\begin{equation*}
  \frac{J_2(\frac{m}{k})}{Y_2(\frac{m}{k})}
     -\frac{J_1(\ee^{R k \pi}\frac{m}{k})}{Y_1(\ee^{R k
         \pi}\frac{m}{k})} = 0
    \quad \longrightarrow \quad  m_n\simeq z_n k \ee^{-R k \pi}\,,
\end{equation*}
where $z_n$ are the positive roots of $J_1$. From $z_0\approx 3.83$ we
find a heavy gravitino solution at the scale of the lightest 
graviton KK mode ($\mathcal{O}(\unit[2]{TeV})$ for $k=\Lambda_{\text{Pl}}$).
Like the heavy gravitons, this gravitino does not have Planck
suppressed interactions but rather a coupling $\lesssim
\unit{TeV}^{-1}$ depending on the localization parameters. To be more precise, let
us calculate the coupling scales 
relevant for
neutralino annihilation (and gravitino decay). The general solutions
for the KK wave functions are (cf.,
e.\,g.,~\cite{Gherghetta:2000kr,Gherghetta:2002nr,Davoudiasl:2000wi})
\begin{align}
  \tilde{G}^{(n)}_1(y)
    &=\ee^{2 R k y}\left[\alpha \,J_2(m_n \,\ee^{R k y}/k)
       + \beta \,Y_2(m_n\, \ee^{R k y}/k) \right]\\
  \tilde{G}^{(n)}_2(y)
    &=\ee^{2 R k y}\left[\alpha \,J_1(m_n \,\ee^{R k y}/k)
       + \beta \,Y_1(m_n\, \ee^{R k y}/k) \right]
     =\frac{\ee^{-R k y}}{R\,m_n} \partial_y \tilde{G}^{(n)}_1\,,
\end{align}
with the canonical normalization condition
\begin{equation*}
  \int\! \dd y\,\ee^{-2 R k y} G^2_i(y)=1\,.
\end{equation*}
With these conventions, the coupling strength to a
vector and a gaugino is
\begin{align}
  \langle\tilde{G}_i f_V f_{\chi_i}\rangle
    &= \frac{1}{\Lambda_{\text{Pl}}} \sqrt{\frac{R}{2 k}}
         \int\!\dd y \ee^{-3/2 R k y}\,\tilde{G}_i f_V f_{\chi_i}\,.
\end{align}
The normalization is chosen such that a gravitino zero mode with unbroken
SUSY would yield
$\langle\tilde{G}_1^{(0)} f_V f_{\chi_1}\rangle=\Lambda_{\text{Pl}}^{-1}$,
$\tilde{G}_2^{(0)}=0$. For $\kappa=0.4$ and $k=\Lambda_{\text{Pl}}$ we obtained
\begin{subequations}
\begin{align}
   \langle\tilde{G}_1 f_Z^{} f_{\chi^0_1}^{(1)}\rangle
      &\approx(\unit[500]{TeV})^{-1}&
   \langle\tilde{G}_2 f_Z^{} f_{\chi^0_2}^{(1)}\rangle
      &\approx(\unit[25]{TeV})^{-1} \label{Zgravitino}\\
   \langle\tilde{G}_1 f_{Z}^{(2)} f_{\chi^0_1}^{(1)}\rangle
      &\approx(\unit[10]{TeV})^{-1}&
   \langle\tilde{G}_2 f_{Z}^{(2)} f_{\chi^0_2}^{(1)}\rangle
      &\approx(\unit[80]{TeV})^{-1} \label{Zgravitino2} \\
   \langle\tilde{G}_1 f_{Z}^{(2)} f^{(2)}_{\chi^{0}_1}\rangle
      &\approx(\unit[1]{TeV})^{-1}&
   \langle\tilde{G}_2 f_{Z}^{(2)} f^{(2)}_{\chi^{0}_2}\rangle
      &\approx(\unit[3]{TeV})^{-1} \label{Zgravitino3}\,.
\end{align}
\end{subequations}
Considering the usual holographic picture, this result is not
surprising, because the heavy gravitino is interpreted as a bound state
with TeV suppressed interactions to other bound states
(\ref{Zgravitino3}), coupling less strongly to lighter, mostly 
elementary particles~(\ref{Zgravitino}) and~(\ref{Zgravitino2}).
With the result in~(\ref{Zgravitino}) for the SM couplings we can
neglect neutralino annihilation
with the gravitino in the $t$-channel in the following discussion.
Still, such a gravitino couples strongly enough to essentially
vanish immediately for a temperature of~$T\ll \unit[2]{TeV}$, long before
nucleosynthesis.

\section{Dark Matter Phenomenology}
\label{sec:dmpheno}

Our aim in this section is to estimate the
relic density of the neutralino LSP we found in the scenarios
discussed above. The analysis of thermal WIMP production and relic
densities generally requires to numerically solve the Boltzmann
equation for the scattering processes involved, but there are
approximate semi-analytic solutions adequate for WIMPs in SM
extensions. In our calculation of the neutralino freezeout
temperature and relic density we follow
\cite{Kong:2005hn,Servant:2002aq}. In the case without
coannihilations\footnote{Even though the neutralinos may be split by
radiative corrections we treat them as one particle with four d.o.f.
rather than having coannihilation between them.} we need the number
of effectively massless degrees of freedom $g^*$, the LSP mass and
the total cross section $\chi_i^0 \chi_j^0 \rightarrow X$ expanded in
the relative velocity~$v$
\begin{equation}
\label{sigmav}
   v \sigma^{tot}_{ij}(v) = a_{ij}+b_{ij}\,v+\mathcal{O}(v^2)\,,
\end{equation}
which we obtain by evaluating the graphs in Fig.~\ref{annihigraphs}.
The cross section~(\ref{sigmav}) can be reexpressed in terms of the
temperature~$T$
\begin{equation}
  \langle v\sigma^{tot}_{ij}\rangle (T)
     \simeq a_{ij}+b_{ij}\,\frac{6 \,T}{m_{\chi^0}}\,.
\end{equation}
The relative freezeout temperature $1/x_f=T_f/m_{\chi^0}$ can be
determined iteratively, and is typically at around $x_f\approx
15\dots 25 $. The final expression for the relic density is then
\begin{equation}
  \Omega h^2(x_f) = \frac{1.04 \times
      \unit[10^9]{GeV}^{-1}}{\Lambda_{\text{Pl}}} \frac{x_f}{\sqrt{g^*}}\frac{1}{a+ 3 b/x_f}\,.
\end{equation}
The effective number of degrees of freedom is taken to be $g^*=86\frac{1}{4}$, 
to which fermions contribute with a weight of $7/8$ from the Fermi distribution.
To give an estimate of the individual contributions
$\chi_i^0 \chi_j^0\rightarrow X_n$, we will add the inverse
densities
\begin{equation}\label{densitysum}
1/\Omega h^2(x_f)=\sum_n {\Omega h^2_n}(x_f)^{-1}\propto \sum_n
a_n\!+3\,b_n/x_f
\end{equation}
for a realistic fixed freezeout temperature, which depends only
weakly on the cross section. The value currently
favored by WMAP data is $1/{\Omega h^2}\approx 9\dots 11$~\cite{Spergel:2006hy}.

\subsection{Neutralino Annihilation Channels}

\begin{figure}
  \begin{center}
    \hfil\\
    \includegraphics{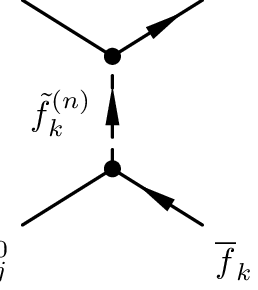}\quad
    \includegraphics{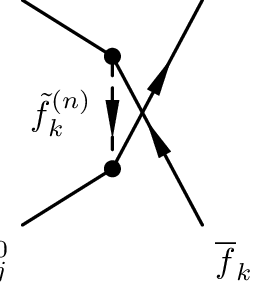}\quad
    \includegraphics{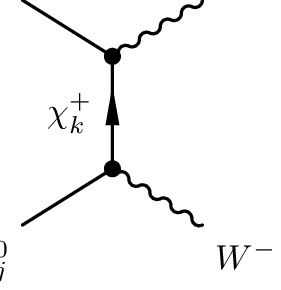}\quad
    \includegraphics{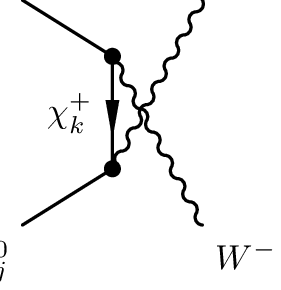}
  \end{center}
  \caption{\label{annihigraphs}%
    Tree level contributions to neutralino annihilation.}
\end{figure}

\begin{figure}
  \begin{center}
    \hfil\\
    \includegraphics{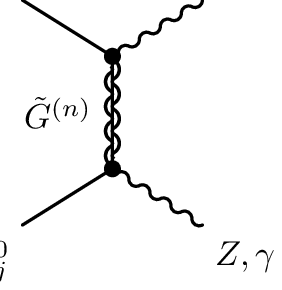}\qquad
    \includegraphics{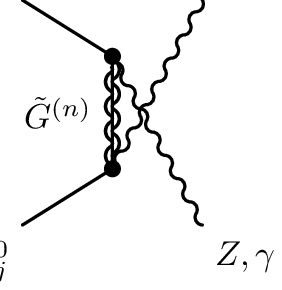}\qquad
    \includegraphics{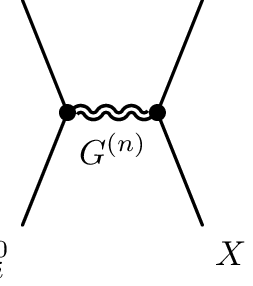}
  \end{center}
  \caption{\label{annihigraphsG}%
    Potential tree level contributions to neutralino annihilation from
    the gravitinos and massive gravitons}
\end{figure}

Depending on the neutralino mass there are several annihilation
channels contributing at tree level (Fig.~\ref{annihigraphs}). The
channels $\chi_i^0\chi_j^0\rightarrow
l\overline{l},\nu\overline{\nu}$ and $\chi_i^0\chi_j^0\rightarrow
d\overline{d},u\overline{u},s\overline{s},c\overline{c},b\overline{b}\to\text{jets}$
are always open, whereas $\chi_i^0\chi_j^0\rightarrow W^+ W^-$ is
open for all parameters in scenario~(\ref{eq:scenario-1}) but only if
$\cos{\theta}\sqrt{1+\kappa}>1$ in scenario~(\ref{eq:scenario-2}).
We first rewrite the neutralinos (charginos) as two
Majorana (Dirac) fermions. To achieve this, we express the 4D action
in terms of the KK coefficient fields $\lambda_1^{1,2(n)}(x)$,
$\lambda_2^{1,2(n)}(x)$, $\lambda_1^{0(n)}(x)$ and
$\lambda_2^{0(n)}(x)$ and group those into 4 spinors
(the KK indices are implicit)
\begin{align}
\chi^{0}_{a}(x)&=\frac{1}{\sqrt{2}}\left(\lambda_1^{0}+\lambda_2^{0}, \overline{\lambda_1^{0}}+\overline{\lambda_2^{0}} \right)
\nonumber \\
\chi^{0}_{b}(x)&=\frac{i}{\sqrt{2}}\left(\lambda_1^{0}-\lambda_2^{0}, \overline{\lambda_2^{0}}-\overline{\lambda_1^{0}} \right)\nonumber  \\
\chi^{+}_{a}(x)&=\frac{1}{\sqrt{2}}\left(\lambda_1^{1}-i\lambda_1^{2}, \overline{\lambda_2^{1}}-i\overline{\lambda_2^{2}} \right)\nonumber  \\
\chi^{+}_{b}(x)&=\frac{1}{\sqrt{2}}\left(\lambda_2^{1}-i\lambda_2^{2},
\overline{\lambda_1^{1}}-i\overline{\lambda_1^{2}} \right)\nonumber
\end{align}
With these definitions, the full 4D chargino-matter-smatter interaction
Lagrangian takes the form
\begin{align}\mathcal{L}&=
h_j^{}\,\overline{\Psi}_i^{}i\Gamma^{}_{ij} (\chi_a^{0}+i\chi_b^{0})+
h_j^{c\dagger}\,\overline{\Psi}_i^{} i\Gamma^{c}_{ij}(\chi_a^{0}-i
\chi_b^{0})+\text{h.c.}\,.
\end{align}
The vertices are given in terms of the projectors, overlap integrals
and quantum numbers by
\begin{align*}
\Gamma_{ij}^{}&=g_5 T_{ij}^{3L}\left[ P^-
\langle f_{\lambda_1^{L3}}f_{\psi_{Li}^{} } f_{h_{Lj}^{}}\rangle- P^+
\langle f_{\lambda_2^{L3}}f_{\psi_{Li}^{c} }f_{h_{Lj}^{}}\rangle\right]\\&+g_5
T_{ij}^{3R}\!\left[ P^- \langle f_{\lambda_1^{R3}}f_{\psi_{Ri}^{}}
f_{h_{Rj}^{}}\rangle- P^+
\langle f_{\lambda_2^{R3}}f_{\psi_{Ri}^{c}} f_{h_{Rj}^{}}\rangle\right]\\
&+\tilde{g}_5 X_{ij}\left[ P^- \langle f_{\lambda_1^{X}}f_{\psi_{Li}^{}}
f_{h_{Lj}^{}}\rangle- P^+ \langle f_{\lambda_2^{X}}f_{\psi_{Li}^{c}}
f_{h_{Lj}^{}}\rangle\right]\\
&+\tilde{g}_5 X_{ij}\left[ P^- \langle f_{\lambda_1^{X}}f_{\psi_{Ri}^{}}
f_{h_{Rj}^{}}\rangle- P^+ \langle f_{\lambda_2^{X}}f_{\psi_{Ri}^{c}}
f_{h_{Rj}^{}}\rangle\right]
\end{align*}
\begin{align*}
\Gamma_{ij}^{c}&=  g_5 T_{ij}^{3L} \left[P^-
\langle f_{\lambda_2^{L3}}f_{\psi_{Li}^{}} f_{h_{Lj}^{c}}\rangle+ P^+
\langle f_{\lambda_1^{L3}}f_{\psi_{Li}^{c}} f_{h_{Lj}^{c}}\rangle \right]\\
&+  g_5 T_{ij}^{3R} \left[P^- \langle f_{\lambda_2^{R3}}f_{\psi_{Ri}^{}}
f_{h_{Rj}^{c}}\rangle+ P^+ \langle f_{\lambda_1^{R3}}f_{\psi_{Ri}^{c}}
f_{h_{Rj}^{c}}\rangle
\right]\\
&+  \tilde{g}_5 X_{ij} \left[P^-
\langle f_{\lambda_2^{X}}f_{\psi_{Li}^{}} f_{h_{Lj}^{c}}\rangle+ P^+
\langle f_{\lambda_1^{X}}f_{\psi_{Li}^{c}} f_{h_{Lj}^{c}}\rangle \right]\\&+
\tilde{g}_5 X_{ij} \left[P^- \langle f_{\lambda_2^{X}}f_{\psi_{Ri}^{}}
f_{h_{Rj}^{c}}\rangle+ P^+ \langle f_{\lambda_1^{X}}f_{\psi_{Ri}^{c}}
f_{h_{Rj}^{c}}\rangle \right]\,.
\end{align*}
In this expression, the brackets $\langle f_{\lambda} f_{\psi}
f_{h}\rangle \equiv \int \!\dd y \sqrt{g}\, f_{\lambda}(y)f_{\psi}(y)
f_{h}(y)$ stand for the 5D overlaps which give us the coupling strengths of a sfermion to the corresponding
matter fermion and a neutralino. The structure is such that the
 ``left handed'' sfermion couples to gauginos
 and matter fermions of the same handedness, the ``right handed'' one
to gauginos and matter fermions of opposite
handedness. The charged current interaction Lagrangian is
\begin{equation}
\mathcal{L}=-\frac{1}{\sqrt{2}} W_m^{+} \left[\overline{\chi^+_a}\,
\Gamma^{m}_a \left(\chi^{0}_a-i\chi^{0}_b\right) +
\overline{\chi^+_b}\, \Gamma^{m}_b
\left(\chi^{0}_a+i\chi^{0}_b\right)\right]+\text{h.c.}
\end{equation}
and the corresponding vertex expressions are
\begin{align*}
\Gamma_a^n&=g_5 \gamma^n \left[P^-(\langle f_{W^{L\pm}}
f_{\lambda^{L\pm}_{2}}
 f_{\lambda^{L3}_{2}}\rangle+\langle f_{W^{R\pm}} f_{\lambda^{R\pm}_{2}}
 f_{\lambda^{R3}_{2}}\rangle)\right. \\&\left.\hspace{4ex}+ P^+(\langle f_{W^{L\pm}} f_{\lambda^{L\pm}_{1}}
 f_{\lambda^{L3}_{1}}\rangle +\langle f_{W^{R \pm}} f_{\lambda^{R\pm}_{1}}
 f_{\lambda^{R3}_{1}}\rangle)\right]\\
\Gamma_b^n&=\Gamma_a^n (P^+\leftrightarrow P^-)\,,
\end{align*}
with the 5D overlap integrals $\langle f_W f_{\lambda_i} f_{\lambda_j} \rangle
\equiv \int \!\dd y \sqrt{g}\,\ee^{R k y}f_{W}(y) f_{\lambda_i}(y)f_{\lambda_j}(y)
$ which now have an additional factor $\ee^{R k y}$ from the inverse
vielbein contained in~$\gamma^\mu$.
The effective coupling constants for the charged current can be
approximated analytically to leading order using $R k \pi\gg 1$ and $m_W,m_\chi\ll k
\ee^{-R k \pi}$ as
\begin{equation}
\label{approxcouplings}
   \langle f_{W^{L\pm}} f_{\lambda^{L\pm}_{2}}
     f_{\lambda^{L3}_{2}}\rangle\simeq-\frac{3}{8 }C
 \qquad
   \langle f_{W^{R\pm}} f_{\lambda^{R\pm}_{2}}
     f_{\lambda^{R3}_{2}}\rangle\simeq-\frac{1}{8 }C\,,
\end{equation}
with
\begin{subequations}
\begin{equation}
  \langle f_{W^{L\pm}} f_{\lambda^{L\pm}_{1}}
    f_{\lambda^{L3}_{1}}\rangle \simeq-C
  \qquad
  \langle f_{W^{R \pm}} f_{\lambda^{R\pm}_{1}}
    f_{\lambda^{R3}_{1}}\rangle\simeq-\frac{1}{48 R k \pi}C
\end{equation}
in scenario~(\ref{eq:scenario-1}) and
\begin{equation}
 \langle f_{W^{L\pm}}
   f_{\lambda^{L\pm}_{1}}
   f_{\lambda^{L3}_{1}}\rangle\simeq-\frac{|\cos(\theta)|}{24 R k \pi}C
 \qquad
 \langle f_{W^{R \pm}} f_{\lambda^{R\pm}_{1}}
    f_{\lambda^{R3}_{1}}\rangle \simeq-\frac{7 |\cos(\theta)|}{48 R k \pi} C
\end{equation}
\end{subequations}
in scenario~(\ref{eq:scenario-2}), respectively, where
\begin{equation}
  C=\frac{1}{\sqrt{(\kappa+1) R \pi}}\,.
\end{equation}
They are accurate to about $5\%$ for $R k
\pi>30$ (cf.~also Fig.~\ref{fannihiW1} below). Note that the absolute
size of the coupling is approximately independent of the free parameters,
\begin{equation}
\label{neutralinocoupling}
  g_5 C\simeq \frac{\sqrt{4 \pi \alpha}}{\sin(\theta_W)}\,,
\end{equation}
and the leading contributions are also independent of the neutralino
mass. For the sfermions with $m_{\widetilde{f}}
> k \ee^{- R k \pi}$ the same expansion is not possible. The Feynman rules
were implemented and the cross sections evaluated using
\texttt{FeynArts/FormCalc}~\cite{Hahn:2000kx}.

\subsection{Results}

\subsubsection{Annihilation into Fermions}

Due to the large KK mass of the sfermions around $-c_R,c_L\approx
0.2 \dots 0.7$ of $m_{\widetilde{f}} > 2 k \ee^{-R k \pi}$, the cross
sections $\chi^0 \chi^0\rightarrow \overline{f}f$ will be suppressed
relative to the annihilation to $W$~pairs if $m_{\chi^0}>m_W$. We find
that the couplings and cross section depend strongly on the
localization of the fermion multiplets, but that a rather extreme
localization towards the TEV brane is necessary to make a
significant contribution to the total annihilation cross section.
This region of parameter space is ruled out due to FCNCs (unless an
additional flavor symmetry is introduced to suppress them) and
proton decay from higher dimensional operators. Therefore we
conclude that in this scenario the tree level annihilation of
neutralinos into fermions is negligible. For example, if one
carries out the calculation for $k=\Lambda_{\text{Pl}}$, $\kappa=.4$,
$x_f=23$ the contribution of the charm quark stays below $1/\Omega
h^2 <0.3$ for $|c_{L,R}|>0.25$ and even $1/\Omega h^2 <0.02$ for
$|c_{L,R}|>0.5$. The electron which must be at $c_L\approx 0.5$ to
get a realistic $S$~parameter, contributes $1/\Omega h^2 <0.01$. One
uncertainty comes from the precise implementation of the third quark
generation. The straightforward generation of $m_b$ and $m_t$ with
boundary terms leads to problems with the $Zb\overline{b}$
coupling\footnote{For a proposed solution
see~\cite{Cacciapaglia:2006gp}.}. Depending on how the $b$ quark is
realized, its coupling to the neutralino could be somewhat enhanced,
but we can still expect that the channel~$\chi^0_i\chi^0_j\rightarrow
b\overline{b}$ does not play a major role for dark matter
annihilation.

\subsubsection{Annihilation into $W$~Pairs}

\begin{figure}
\begin{center}
\includegraphics[width=.6\linewidth]{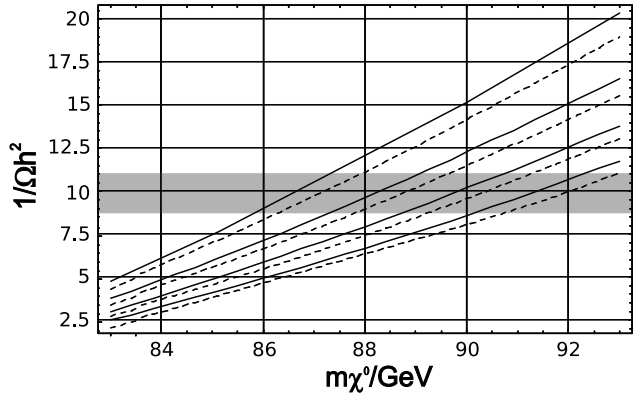}
\caption{\label{fannihiW1}%
  The $m_{\chi^0}$-dependence of contributions to $1/\Omega h^2$ for
  annihilation into $W$ pairs for $R k \pi\simeq37.5$,
  $k=\Lambda_{\text{Pl}}$. The parameter is (from topmost)
  $\kappa=0.4,0.6,0.8,1.0$, corresponding to
  $m_{\chi^+}=95,102,108,114$ GeV. For this process we use $\alpha=
  1/128$. The dashed lines are obtained using the approximations
  in~(\ref{approxcouplings})-(\ref{neutralinocoupling}). The shaded
  area represents the region allowed by the WMAP three year data.  }
\end{center}
\end{figure}

If $m_{\chi^0}>m_W$, the neutralino LSP can annihilate into two $W$s
at tree level, and the interaction is essentially of weak strength
(\ref{neutralinocoupling}). In scenario~(\ref{eq:scenario-1}) the annihilation
cross section turns out to be too large, leading to a small relic density
which would require another mechanism for Dark Matter production.
In scenario~(\ref{eq:scenario-2}) the largest coupling constants are strongly
suppressed, and the relic density is in a realistic range
(Fig.~\ref{fannihiW1}).
A neutralino LSP with $m_{\chi^0}\gtrsim \unit[86]{GeV}$ is favored.

\section{Conclusions}
\label{sec:conclusions}

We have constructed a minimal supersymmetric extension of warped
higgsless models. The absence of SUSY on the UV brane leads to a
tree level spectrum in which the lightest neutralinos are naturally the
lightest superpartners and the lightest charginos the NLSPs.
Experimental constraints force us to shift the chargino mass
above~$\unit[95]{GeV}$, with~$m_{\chi^0}=0\,\ldots\,m_{\chi^+}$
depending on the mixing angle. All color charged superpartners and the
sfermions are at the KK scale~$\lesssim \mathcal{O}(\unit{TeV})$.

The lightest electroweak gauge scalars~$\Sigma^\pm$, $\Sigma^0$ have
tree level masses at~$\sqrt{2}\,m_{\chi^+}$, but are expected to
receive significant radiative corrections. Their Higgs-like masses and
couplings to matter make them an interesting object for
searches at the LHC, where they could be distinguished from the Higgs
by the absence of Higgsstrahlung and vector boson fusion processes.
These issues will be treated in an upcoming
publication, along with a more detailed analysis of LEP precision
observables.

Another aspect to be investigated is tree level unitarity of the
model which generally requires vector boson KK modes lighter than
the ones found in our minimal scenario, a feature which can for
example be achieved by including additional boundary terms.

For $m_{\chi^0}>m_W$, the dominant channel for LSP annihilation is
into $W$ pairs, while the annihilation into SM fermions does not
contribute significantly unless they are highly localized towards
the TeV brane. If the left-handed neutralino is set up to be
localized mostly in the $SU(2)_R$ triplet, the LSP relic density is
in agreement with current observations for a natural range of
parameters where $m_W<m_{\chi^0}<m_{\chi^+}$, making SUSY in the
EWSB sector a promising source for the observed nonbaryonic dark
matter if Higgsless symmetry breaking turns out to be realized in
nature.

\section*{Acknowledgments}
A.\,K.~thanks Alex Pomarol for useful discussions
and the hospitality extended to him at IFAE. The authors thank Hitoshi
Murayama for useful discussions on EWSB, AdS/CFT and supersymmetry.
The authors also thank Dominik Els\"asser for useful discussions on
the dark matter relic density.

This research is supported by Deutsche Forschungsgemeinschaft through
the Research Training Group 1147 \textit{Theoretical Astrophysics and
Particle Physics} and by Bundesministerium f\"ur Bildung und
Forschung Germany, grant 05HT6WWA.

\appendix

\section{Conventions}
\label{appendix}
We use the flat metric convention $\eta=\text{diag}(+,-,-,-,-)$ and
the corresponding Dirac matrices
\begin{align}
\gamma^m=-\left(\begin{array}{cc} 0&\sigma^m\\
\overline{\sigma}^m&0\end{array}\right)~~\gamma^{\bar{5}}=\left(\begin{array}{cc} i&0\\
0&-i\end{array}\right)\,,
\end{align}
where $\sigma^0=\overline{\sigma}^0=-1$ and $-\overline{\sigma}^i=\sigma^i$ are the Pauli
matrices. We define the projectors
\begin{align}P^+=\frac{1}{2}(-i\gamma^{\bar{5}}+1)\quad
P^-=\frac{1}{2}(-i\gamma^{\bar{5}}-1)
\end{align}
on the 4D ``left handed'' and ``right handed'' component respectively.
For Dirac spinors $\Psi^T=(\eta_\alpha,\overline{\chi}^{\dot{\alpha}})$ we
define $\overline{\Psi}=\Psi^\dagger
\gamma^{\overline{0}}=(\chi^{\alpha},\overline{\eta}_{\dot{\alpha}})$.

Two coordinate systems are commonly used. The ``proper distance''
coordinates have $x^M=(x^\mu,y)$ and $y\in[0,\pi]$. They are related
to the ``conformal'' coordinates with $x^M=(x^\mu,z)$,
$z\in[k^{-1},\Lambda_{\text{IR}}^{-1}]$ through
\begin{align}
z=k^{-1} \ee^{R k y}\qquad \Lambda_{\text{IR}}=k \ee^{- R k \pi}\,.
\end{align}
The Dirac matrices read in proper distance coordinates
\begin{subequations}
\begin{align}
\gamma^\mu&=-\delta^\mu_m \ee^{R k y}\left(\begin{array}{cc} 0&\sigma^m\\
\overline{\sigma}^m&0\end{array}\right)\\\gamma^{5}&=\frac{1}{R}\left(\begin{array}{cc} i&0\\
0&-i\end{array}\right)\\
\gamma_\mu&=\eta_{\mu \nu} \ee^{-2 R k y}\gamma^\nu\\
\gamma_5&=-R^2 \gamma^5 \\
\end{align}
\end{subequations}
and in conformal coordinates
\begin{subequations}
\begin{align}
\gamma^\mu&=-\delta^\mu_m k z\left(\begin{array}{cc} 0&\sigma^m\\
\overline{\sigma}^m&0\end{array}\right)\\\gamma^{5}&=k z\left(\begin{array}{cc} i&0\\
0&-i\end{array}\right)\\
\gamma_\mu&=\eta_{\mu \nu}\frac{1}{k^2 z^2} \gamma^\nu\\
\gamma_5&=-\frac{1}{k^2 z^2} \gamma^5\,.
\end{align}
\end{subequations}

\section{Boundary Conditions and KK Decomposition}
\label{sec:gaugebckk}
\subsubsection*{Boundary Conditions}
The kinetic terms that we obtain straight from the superfield action and
the BCs take the canonical form after the field
redefinitions
\begin{equation}
\label{redef}
\begin{aligned}
 &\psi \longrightarrow \ee^{-\frac{1}{2}Rky}\psi&&
A_\mu\longrightarrow g A_\mu &&
\lambda_1 \longrightarrow g \ee^{-\frac{3}{2}Rky}\lambda_1\\
&A_5 \longrightarrow g R A_5 && \lambda_2 \longrightarrow -i g R
\ee^{-\frac{1}{2}Rky}\lambda_2 && \Sigma \longrightarrow  g R
\Sigma
\end{aligned}
\end{equation}
The standard coupling constants are recovered after taking
$g_5\equiv g/2$. Before we specify the KK decomposition for the
fields, we employ an $R_\xi$ type gauge fixing term
\begin{align}
\label{eqgaugefix} S_{gf}=-\int\!\dd^5 x\,
\frac{R}{2\xi}\left(\eta^{mn}\partial_m A_n- \xi \frac{\ee^{-2 R k
y}}{R}(\partial_y-2 R k)  A_5\right)^2\,.
\end{align}
The BCs in~(\ref{superbcs}) now give the KK modes
$A_5^{(n)}$ an unphysical mass $\sqrt{\xi} m_n$ if the corresponding
massive gauge boson mode has a mass $m_n$. This can be seen as
follows: With~(\ref{eqgaugefix}) and~(\ref{eqgaugeaction}) the KK
equations of motion are
\begin{align*}
m_n^2 A^{(n)}_\mu &= \ee^{-2 R k y}(-\partial_y^2/R^2 +2 k \partial_y/R)A^{(n)}_\mu\\
m_n^2 A^{(n)}_5 &= \ee^{-2 R k y}(-\partial_y^2/R^2 +4 k
\partial_y/R-4 k^2)A_5^{(n)}\,.
\end{align*}
Taking the derivative of the first equation, we find
\begin{align}
m_n^2 \partial_y A^{(n)}_\mu &= \ee^{-2 R k y}(-\partial_y^2/R^2 +4 k
\partial_y/R-4 k^2)
\partial_y A^{(n)}_\mu \label{gaugeobcrule}\,.
\end{align}
Since $\partial_y A_\mu$ satisfies the same equation of motion as
$A_5$, the correct choice of BCs immediately follows:
\begin{equation}
\begin{aligned}
\partial_y A_\mu|_{\partial}=0&\Longrightarrow A_5|_{\partial}=0 \\
A_\mu|_{\partial}=0&\stackrel{m_n\neq0}{\Longrightarrow} (\partial_y -2
R k)A_5|_{\partial}=0\,.
\end{aligned}
\end{equation}
Now, there is a mode of $A_5$ for every massive mode of $A_\mu$. A
similar reasoning applies to the fermions. Consider the action for
any gaugino,
\begin{align}
\int \!\dd^5x\:R \ee^{-3 R k y} (-i \lambda_i \sigma^n \partial_n
\overline{\lambda}_i)+\int \!\dd^5x \: R \ee^{-4 R k y}\left[-\lambda_2
\frac{\partial_y - 3/2 R k}{R}\lambda_1 +c.c. \right]\,.
\end{align}
The KK decomposition for these coupled differential equations
diagonalizes the KK Dirac mass if
\begin{equation}
\label{gauginobcrule}
\begin{aligned}
(\partial_y -3/2 R k)\lambda_1=0&\Rightarrow \lambda_2=0 \\
\lambda_1=0&\Rightarrow (\partial_y -5/2 R k)\lambda_2=0
\end{aligned}
\end{equation}
This choice is again related to (\ref{gaugeobcrule}) by
(\ref{eq:susypar}).

\subsubsection*{Brane Kinetic Terms}
We will eventually introduce a gauge invariant UV brane kinetic term
for the $SU(2)_L$ gauge bosons to split the $W^\pm$ and the
charginos. Localized gauge kinetic terms modify the BCs,
the definition of the scalar product and of the coupling
constants.
For example, such a term for the $SU(2)_L$ vectors
\begin{equation}
\label{su2bkt}
S\rightarrow S+\kappa\int\!\dd^5x\,\delta(y)\pi R\,
\left(-\frac{1}{4}\eta^{mo}\eta^{np}L^a_{mn}L^a_{op}\right)\,,
\end{equation}
leads to the canonical normalization conditions for the $W^\pm$, $Z$
and photon
\begin{gather}
\int\!\!\dd y\, R\Big(f^{L12(n)}(y)^2+f^{R12(n)}(y)^2\Big)+\pi R \kappa f^{L12(n)}(0)^2=1\\
\int\!\!\dd y\, R\Big(f^{L3(n)}(y)^2+f^{R3(n)}(y)^2+f^{X(n)}(y)^2\Big)+\pi R \kappa f^{L3(n)}(0)^2=1\\
\pi R\left((2+\kappa)\left(\frac{a_0}{g_5}\right)^2
+\left(\frac{a_0}{\tilde{g}_5}\right)^2\right)=1\,.
\end{gather}
The Neumann BC $\partial_y A^L_\mu(0)=0$ becomes
\begin{equation*}
\left(\partial_y+m^2 \pi R^2 \kappa \right) A^L_\mu(0)=0\,.
\end{equation*}
The dimensionless constant $\kappa$ is naturally $\simeq\mathcal{O}(1)$.

\subsubsection*{KK Decomposition}

The 5D fields are split into a coefficient bearing the 4D
dependence, and the KK wavefunctions
\begin{align}
\phi(x,y)=\sum_n \phi^{(n)}(x) f^{(n)}(y)\,.
\end{align}
The canonical normalization of the wavefunctions is $\int_0^\pi\!\dd y \,R f(y)^2=1$ in
flat space, but depends on the field in warped space because of the
factor $\sqrt{g}$ and the different number of vielbeins $e_\mu^n$.
For vectors, spinors and scalars respectively it is
\begin{align}
\int\!\!\dd y R f_V^2(y) =1,\quad\int\!\!\dd y R \ee^{-3 R k y}
f_\Psi^2(y) =1, \quad\int\!\!\dd y R \ee^{-2 R k y} f_\Sigma^2(y) =1\,.
\end{align}
Any BC mixing two fields on the boundaries
causes the two fields to belong to the same 4D KK tower,
e.g.~$A_\mu^{L(n)}\!(x)f^{L(n)}(y)=A_\mu^{R(n)}\!(x)f^{R(n)}(y)$ requires
$A_\mu^{L(n)}(x)=A_\mu^{R(n)}(x)$ to have the free action diagonal
in the KK modes. Canonical normalization then means that the sum of
the normalizations of all fields coupled in this manner should be
unity. 

\end{document}